\documentclass[%
reprint,
groupedaddress,
twocolumn,
amsmath,amssymb,aps,showpacs,
aps,prb,
floatfix,
]{revtex4-1}

\usepackage{graphicx,amssymb,amsmath,epsfig}
\usepackage{color}
\usepackage[latin1]{inputenc}

\newcommand{\wmk}{Wm$^{-1}$K$^{-1}$}

\begin{document}

\title{Divergence of the Thermal Conductivity in Uniaxially Strained Graphene}
\author{Luiz Felipe C. Pereira}
\email{pereira@mpip-mainz.mpg.de}
\affiliation{Max Planck Institute for Polymer Research, Ackermannweg 10, D-55128 Mainz, Germany}
\author{Davide Donadio}
\email{donadio@mpip-mainz.mpg.de}
\affiliation{Max Planck Institute for Polymer Research, Ackermannweg 10, D-55128 Mainz, Germany}

\date{\today}

\begin{abstract}
We investigate the effect of strain and isotopic disorder on thermal transport in suspended graphene by equilibrium molecular dynamics simulations.
We show that the thermal conductivity of unstrained graphene, calculated from the fluctuations of the heat current at equilibrium is finite and converges with size at finite temperature. In contrast, the thermal conductivity of strained graphene diverges logarithmically with the size of the models, when strain exceeds a relatively large threshold value of 2\%.
An analysis of phonon populations and lifetimes explains the divergence of the thermal conductivity as a consequence of changes in the occupation of low-frequency out-of-plane phonons and an increase in their lifetimes due to strain.
\end{abstract}

\pacs{65.80.Ck, 63.22.Rc, 05.60.Cd}

\keywords{Graphene, thermal conductivity, strain, low-dimensional materials, molecular dynamics}

\maketitle

\section{Introduction}

The combination of light weight, strong covalent bonds, and low dimensionality gives carbon nanostructures, such as graphene and nanotubes, superior mechanical and thermal properties~\cite{Balandin2011}, making them interesting candidate materials for thermal management~\cite{Ghosh2008} and  phononics applications~\cite{Yang2009, Hu2009a, Li2012b}.
Extremely high, possibly divergent, thermal conductivity ($\kappa$) of a two-dimensional (2D) phonon gas was predicted by Klemens and Pedraza ten years before single layer graphene was isolated for the first time~\cite{Klemens1994}.
The first measurements in suspended graphene at about room temperature confirmed Klemens and Pedraza's predictions, reporting  values of $\kappa$ in the range $3000 - 5800$ \wmk ~\cite{Balandin2008,Ghosh2008}.
Later experiments found $\kappa \approx 2500$ \wmk at $350$ K, and  $\approx 1400$ \wmk at $500$ K~\cite{Cai2010a}. Recent measurements on suspended graphene in vacuum yield $\kappa$ in the range $2600 - 3100$ \wmk at $350$ K~\cite{Chen2011a}. 

In spite of the efforts to refine these measurements, an accurate determination of  $\kappa$ remains a tough experimental challenge~\cite{Balandin2011}. A fundamental reason for such difficulty is that heat transport in graphene is very sensitive to defects and experimental conditions.
For example, when graphene is supported on a substrate $\kappa$ is reduced to $\approx 600$ \wmk at room temperature due to  phonon scattering from the substrate~\cite{Seol2010}.
$\kappa$ can vary by as much as 50$\%$ at room temperature as a function of the isotopic composition of graphene~\cite{Chen2012}, and  was found to be sensitive also to the lateral dimension of measured patches~\cite{Nika2012} and 
to the presence of wrinkles, which may lower $\kappa$ by $\sim 30\%$~\cite{Chen2012a}.
This high sensitivity can be advantageous because it offers the possibility to manipulate $\kappa$ in graphene-based devices either by tuning the concentration of defects and mass disorder during growth or by imposing controlled external conditions.
High and tunable thermal conductivity for a single sheet of atoms opens up the possibility of application in a range of thermal management devices, from high-power electronics~\cite{Yan2012} all the way to thermoelectric applications~\cite{Kim2012}.

In pristine graphene at room temperature, heat is conducted almost exclusively by phonons~\cite{Ghosh2008}, so we can focus on lattice thermal conductivity.
Both lattice dynamics (LD) calculations~\cite{Lindsay2010,Lindsay2010b,Singh2011,Bonini2012} and molecular dynamics (MD) studies~\cite{Chen2012,Guo2009,Evans2010,Thomas2010} indicate that $\kappa$ of suspended graphene converges with system size, in contrast with ideal 2D models,  for which $\kappa$ diverges logarithmically~\cite{Lippi2000,Lepri2003,Wang2012}. Flexural phonons (out-of-plane vibrational modes) play a decisive role both as heat carriers and as scatterers.
Nevertheless, some of these works report very diverse numerical results, stemming from the use of different methods relying on different approximations, and from the choice of various interatomic potentials. 
Theoretical studies suggest that it is possible to control $\kappa$ in graphene by applying mechanical (tensile) strain, however simulations have given contradictory results. 
Recent {\sl ab initio} LD calculations showed that, while $\kappa$ is finite for unstrained suspended graphene, it diverges when tensile strain is applied~\cite{Bonini2012}. In contrast, MD results point in the opposite direction, indicating a reduction of $\kappa$ upon strain~\cite{Li2010}. 
Even though discrepancies between LD and MD results are expected, as the two methods rely upon different approximations~\cite{Turney2009a,He2012a}, it is unusual to get such differences in  trends. In fact, in LD calculations anharmonic interactions are usually truncated at the first order, while in MD simulations quantum effects cannot be taken into account, so phonon populations obey to classical statistics. Both approximations conspire to make $\kappa_{LD}$ larger than $\kappa_{MD}$.

In this work we report the results of equilibrium molecular dynamics (EMD) simulations of heat transport in suspended graphene as a function of strain and isotopic mass disorder. 
We begin by investigating size convergence of $\kappa$ in isotopically pure unstrained graphene, and then we study the effects of mechanical strain and isotopic disorder. Our goal is to verify whether the  divergence, predicted by LD calculations on isotropically strained graphene, also occurs upon uniaxial strain at finite temperature. We also probe how the combination of strain and mass disorder affects $\kappa$.  A microscopic interpretation of the results is provided in terms of phonon populations and lifetimes computed at finite temperature.

\section{Methods}

We compute the thermal conductivity of graphene by EMD simulations in models with periodic boundary conditions. We use the Tersoff empirical potential~\cite{Tersoff1988} recently re-parametrized to accurately reproduce the vibrational properties of carbon nanostructures~\cite{Lindsay2010}. 
Anharmonic LD calculations employing this set of parameters result in a thermal conductivity $\kappa = 3500$ \wmk for a $10$ $\mu$m graphene flake at $T=300$ K, well within the range of experimental measurements. 
MD production runs are performed in the microcanonical ensemble,  starting from initial configurations equilibrated at the target temperature~\cite{Shinoda2004}. Temperatures between 300 and 1000 K are considered. 
The equations of motion are integrated with a 1~fs time step.
The cell parameters are optimized at the simulation temperature to achieve zero stress in the  (zig-zag) direction, perpendicular to the strained one.

Following linear response theory, $\kappa$ is computed from the integral of the autocorrelation function of the heat flux $\mathbf{J}(t)$ in a microcanonical simulation, according to the Green-Kubo formula \cite{Green1954, Kubo1957}
\begin{equation}
\kappa_{\alpha \beta} =  \frac{1}{k_BT^2}\lim_{t\to\infty} \lim_{V\to\infty} \frac{1}{V} \int_0^t \langle J_{\alpha}(t') J_{\beta}(0) \rangle dt',
\label{eq:greenkubo}
\end{equation}
where $k_B$ is Boltzmann's constant, $T$ is the temperature and $V$ the volume, which is here defined as the surface area of the graphene foil times a nominal thickness of $3.35$ \AA. 
In practice, $\kappa$ is taken as the stationary value of Eq. (\ref{eq:greenkubo}) before it drifts due to accumulated statistical noise.
Although $\kappa$ is in general a tensor, the hexagonal symmetry of graphene yields $\kappa_{xx}=\kappa_{yy}=\kappa$ and $\kappa_{xy}=0$.  
The limits to infinite time and infinite volume in Eq. (\ref{eq:greenkubo}) indicate that size convergence and phase space sampling have to be carefully considered. This aspect is particularly important for low-dimensional systems, for which transport coefficients usually diverge \cite{Lippi2000,Lepri2003,Lepri2005,Wang2012}. Therefore, investigating size and time convergence is not merely a technical aspect, but it brings important physical insight.
In order to effectively sample the phase space and achieve statistical accuracy in evaluating Eq. (\ref{eq:greenkubo}), each reported value of $\kappa$ is obtained by averaging over at least $20$ independent simulations of at least $60$ ns.

\section{Results and discussion}

\subsection{Thermal conductivity of unstrained suspended graphene}
\begin{figure}
\center
\includegraphics[width=0.85\linewidth]{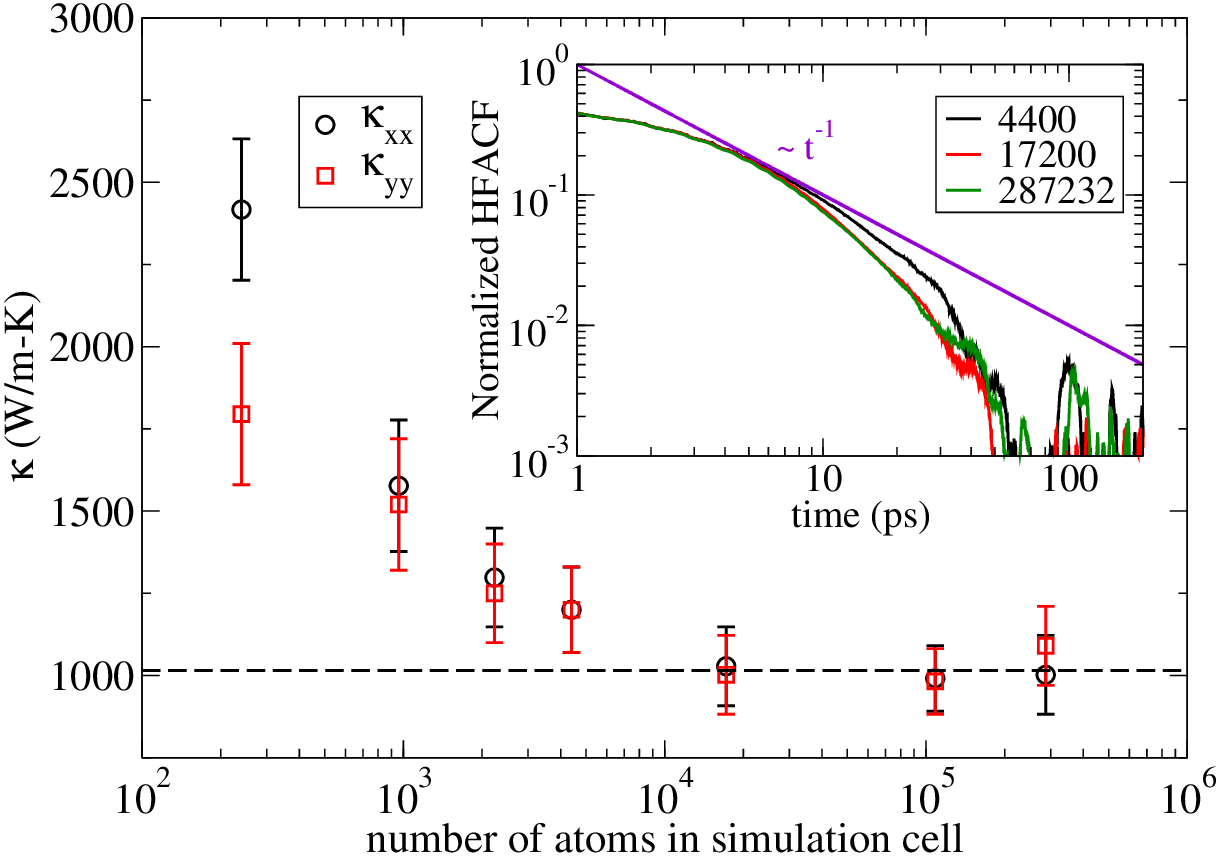}
\caption{(Color online) Thermal conductivity of graphene as a function of the number of atoms in the simulation cell. Converged value $\kappa = 1015 \pm 120$ \wmk is achieved for a cell with $17200$ atoms. Inset: time decay of the normalized heat flux autocorrelation functions (HFACF) for $4400$, $17200$, and $287232$ atom cells. All HFACF decay faster than $1/t$.}  
\label{fig:gk1}
\end{figure}

To check size convergence we perform simulations with approximately square supercells of increasing size. We consider systems made of from $240$ to $3\cdot 10^5$  atoms. The smallest supercell is $26 \times 25$ \AA$^2$ and the largest one $880 \times 877$ \AA$^2$.
Fig.~\ref{fig:gk1} shows the calculated thermal conductivity as a function of the number of atoms in the simulation cell. 
The anisotropy between the in-plane thermal conductivities seen for the smallest cell ($240$ atoms) is due to an uneven and insufficient sampling of the vibrational modes in the two directions. 
As the cell size is increased a better sampling  is achieved, and the anisotropy vanishes.
We find that a $216 \times 214$ \AA$^2$ simulation cell, containing $17200$ atoms, is required to obtain a converged value $\kappa = 1015 \pm 120$ \wmk. Our estimate of $\kappa$ is lower than the values reported in recent works, in which smaller systems were simulated~\cite{Bagri2011,Haskins2011a, Zhang2011}. 
The inset in Fig. \ref{fig:gk1} displays the normalized heat flux autocorrelation functions (HFACF) for several simulations with different cell sizes, 
showing that in all cases the time decay is faster than $1/t$, which guarantees convergence of Eq. (\ref{eq:greenkubo}). 
We can conclude that $\kappa$ of unstrained graphene at finite temperature is finite and converges with size, confirming the prediction of former {\sl ab initio} LD calculations~\cite{Bonini2012}. 

Classical calculations of $\kappa$ far below the Debye temperature ($\Theta_D\sim 2000 K$ for graphene) may yield large differences with respect to calculations taking into account the proper quantum statistics for phonons. 
Two effects contribute to such differences, yet in opposite directions: classical calculations give shorter phonon lifetimes than quantum calculations, but classical phonon heat capacities are always larger than quantum ones.  
LD calculations showed that in graphene at room temperature these two effects compensate to the point that classical $\kappa$ underestimates quantum $\kappa$ only by about 10$\%$~\cite{Singh2011}.

\begin{figure}
\center
\includegraphics[width=0.85\linewidth]{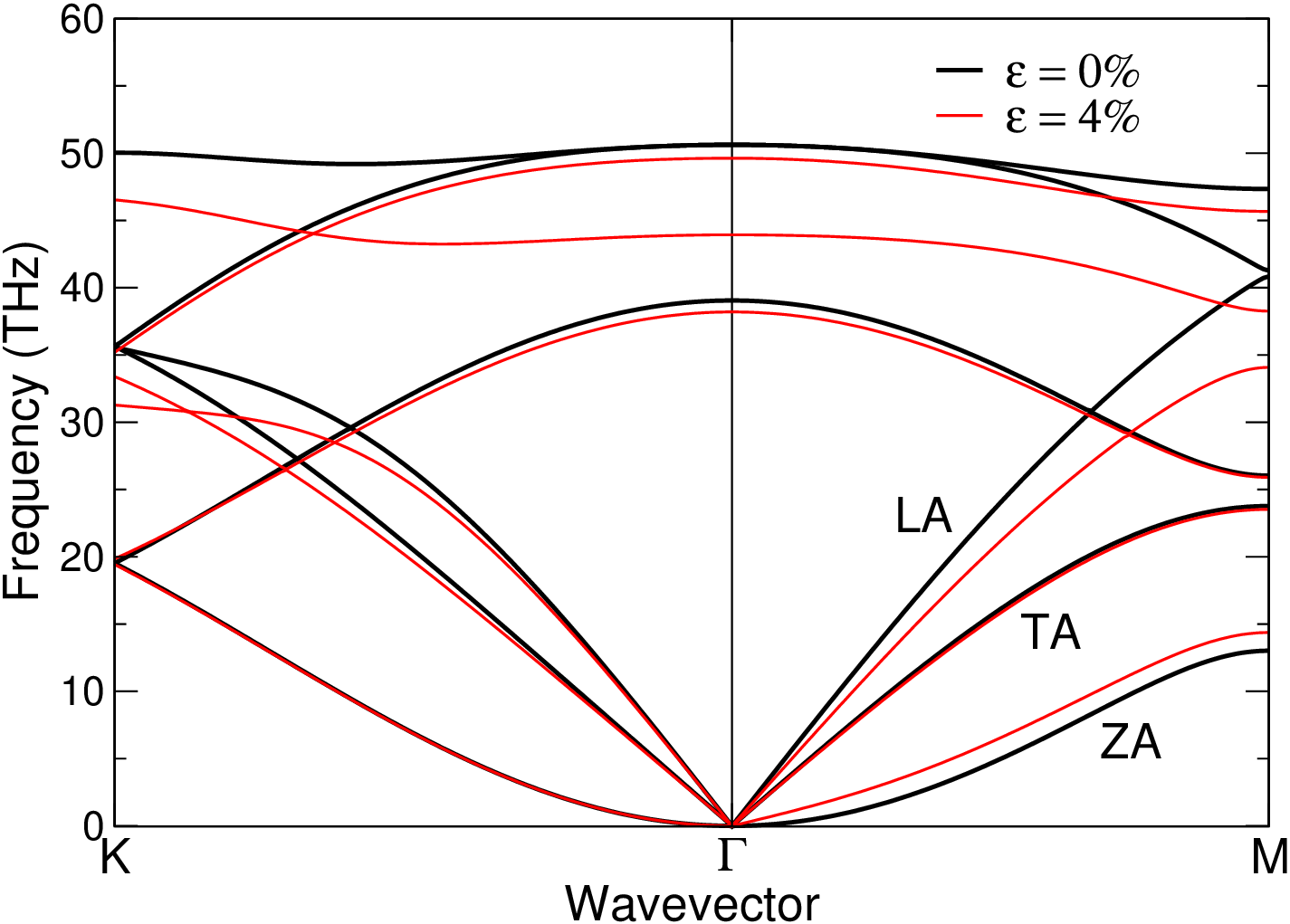}
\caption{(Color online) Dispersion relations of unstrained graphene (black) and of graphene under uniaxial strain of 4$\%$ in the zig-zag direction (red). Dispersion curves are computed in the strained armchair direction ($\Gamma$-M) and in the stress-free zigzag direction ($\Gamma$-K). Wavevectors are in units of the inverse lattice constant $2\pi /a$, rescaled according to applied strain.}
\label{fig:dispersions}
\end{figure}
\begin{figure}
\center
\includegraphics[width=0.85\linewidth]{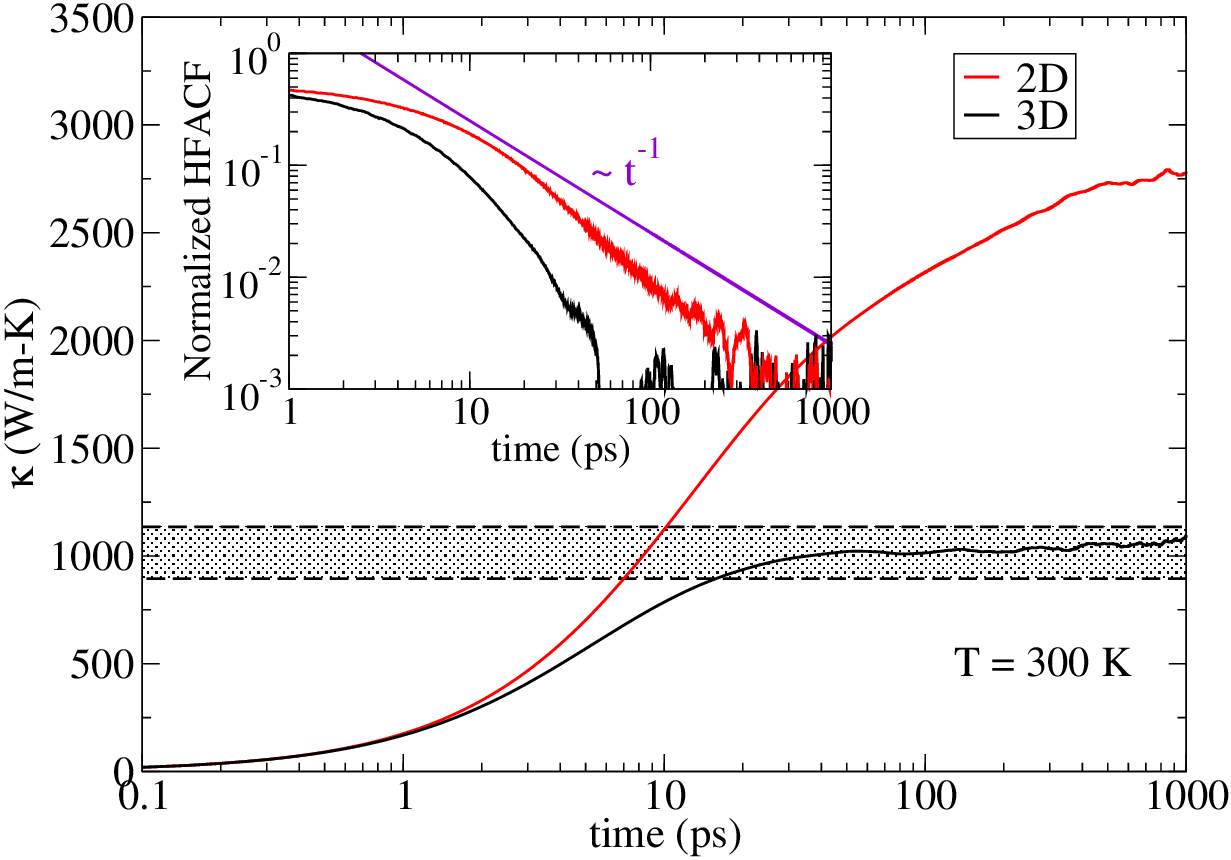}
\caption{(Color online) Average thermal conductivity of graphene in the presence of flexural vibrational modes (3D), and in the absence of such modes (2D), calculated from the Green-Kubo relation. The dashed horizontal lines indicate the uncertainty for the case of 3D graphene. The inset shows the  normalized heat flux autocorrelation functions, which decay like $1/t$ for the 2D case, yielding $\log(t)$ divergence of $\kappa$.}
\label{fig:k.vs.time.2D3D}
\end{figure}

The observed size convergence of $\kappa$ from above, provides an insight into the contribution of flexural phonons to thermal transport in graphene. 
Convergence trends can be interpreted by referring to the dispersion relations of phonons in graphene, computed in the harmonic approximation by diagonalizing the dynamical matrix~\cite{Dove1993} (Fig.~\ref{fig:dispersions}). 
Accurate MD calculations of the thermal conductivity of graphene or carbon nanotubes require a converged sampling of the low-frequency acoustic  flexural (ZA) modes~\cite{Donadio2007}. Whereas the contribution to $\kappa$ of in-plane acoustic modes converges relatively fast, good sampling of the ZA modes is achieved only for large simulation cells because of their quadratic dispersion relation near the $\Gamma$ point.
ZA modes are expected to provide a significant contribution to heat transport, and have been identified as the majority heat carriers \cite{Lindsay2010b}. However, close to the $\Gamma$ point their group velocity vanishes and their main role in thermal transport is to scatter other heat carriers. 
Our convergence trends indeed show that the overall effect of low-frequency ZA modes is to lower the in-plane thermal conductivity of graphene. In fact, performing simulations on `2D graphene', i.e., a graphene sheet in which atoms move only in plane, we observe logarithmic divergence of $\kappa(t)$, as shown in Fig.~\ref{fig:k.vs.time.2D3D}, in accordance with theoretical and numerical studies on 2D model systems \cite{Lippi2000,Lepri2003,Wang2012}.
Our results demonstrate the dual role of ZA modes, which is to provide an important reservoir of heat carriers, as well as the main scattering channel that prevents the divergence of $\kappa$ ~\cite{Bonini2012,Lindsay2010b}.

\subsection{Thermal conductivity of strained graphene}
Strain affects the vibrational properties of materials, as it modifies phonon dispersion relations. Speed of sound, frequency range, scattering rates, and therefore thermal conductivity, are all altered.
We apply uniaxial tensile strain along the armchair direction and relax the simulation cell to achieve zero stress in the perpendicular (zigzag) direction. The dispersion relations of strained graphene (strain $\varepsilon = 4\%$) are compared to the unstrained ones in Fig.~\ref{fig:dispersions}.
In the low-frequency range the most significant changes is the
linearization of the dispersion relation of the ZA mode along the strain axis. In the direction
perpendicular to the strain axis the ZA branch remains unchanged. This implies non-vanishing
group velocity for the ZA modes propagating along the strain axis. In addition, in-plane acoustic
modes are slightly softened in both directions. In the high-frequency range the degeneracy of the
zone center optical phonon is broken, in accordance with Raman measurements.\cite{Mohiuddin2009}

\begin{figure}
\center
(a)\\
\includegraphics[width=0.80\linewidth]{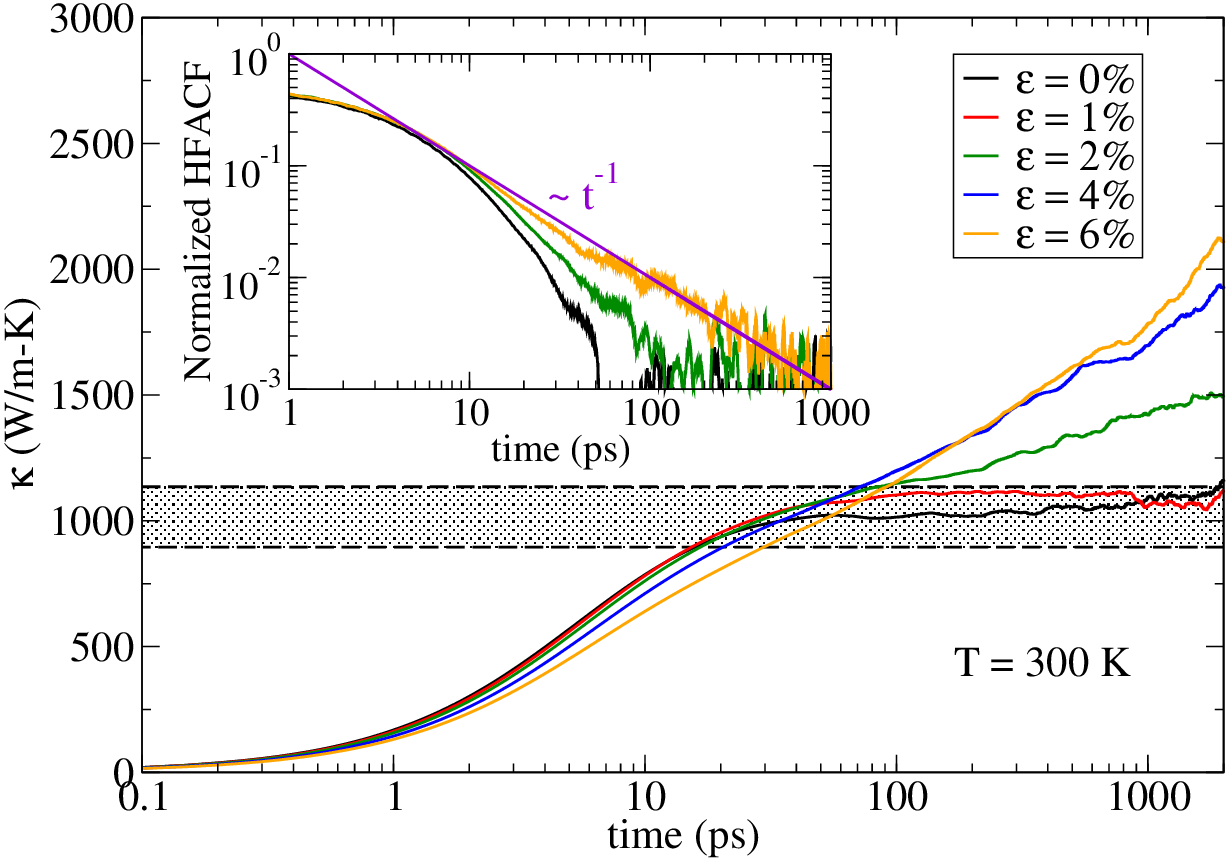}\\
(b)\\
\includegraphics[width=0.80\linewidth]{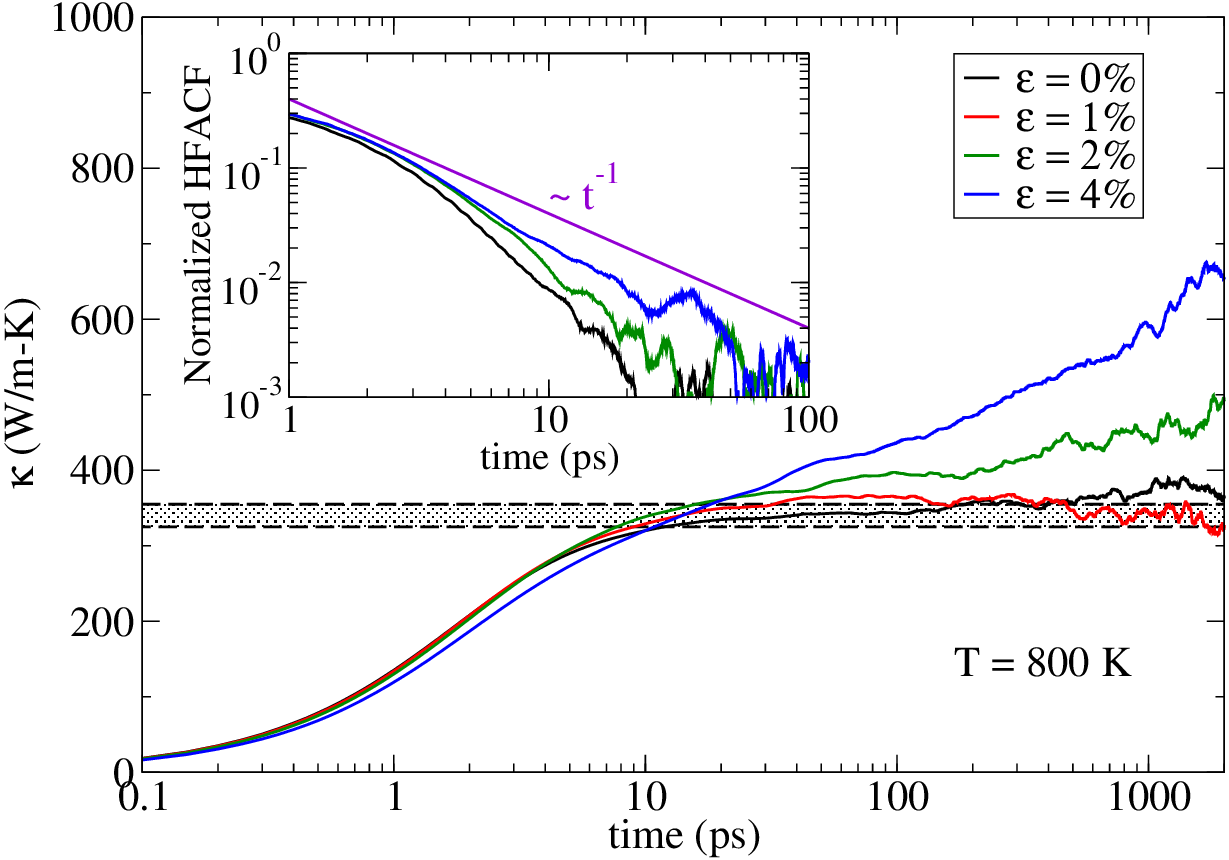}
\caption{(Color online) Thermal conductivity of strained graphene at (a) $300$ K and (b) $800$ K, calculated from Eq. (\ref{eq:greenkubo}). The dashed horizontal lines indicate the average $\kappa$ for unstrained graphene. The inset shows the time decay of the normalized HFACF. For high strain, the HFACF decays as $1/t$ such that $\kappa$ diverges as $\log(t)$ as observed in the main panel.}
\label{fig:kappa.strain}
\end{figure}

Fig.~\ref{fig:kappa.strain} shows the thermal conductivity of graphene as a function of time, calculated as the argument of the time limit in Eq. (\ref{eq:greenkubo}), along the strained direction for strain up to $6\%$. 
For small strain ($\varepsilon=1\%$), the thermal conductivity in the strained direction still converges, yet to a larger value than in the unstrained case. 
As strain is increased ($\varepsilon \geq 2\%$ in the figure), the thermal conductivity along the strained direction tends to diverge.
The same behavior is observed for a larger simulation cell containing more than $10^5$ atoms.
The inset in Fig.~\ref{fig:kappa.strain} displays the time decay of the respective HFACF. For $\varepsilon < 2\%$, the HFACF decays faster than $1/t$. However, at larger strain the HFACF decays as $1/t$, such that $\kappa$ diverges as $\log(t)$, following the standard behavior of transport coefficients in 2D systems~\cite{Lepri2003}. 
Meanwhile, the thermal conductivity in the stress-free direction does not diverge~\footnote{See Supplemental Material at [URL will be inserted by  publisher] for thermal conductivity parallel and perpendicular to strain  direction, thermal conductivity of isotropically strained graphene at $300$ K, VDOS of strained and unstrained graphene, and a comparison of phonon populations in the presence of strain and isotopic mass disorder.}. In fact, the thermal conductivity perpendicular to the strain direction is slightly reduced, due to a mild softening of the in-plane acoustic modes as shown in Fig.~\ref{fig:dispersions}. 

It is important to point out that a logarithmic divergence of $\kappa$ with size in finite 2D model systems under stationary non-equilibrium conditions implies a logarithmic divergence of $\kappa$ as a function of time in periodic systems at equilibrium and {\it vice versa}~\cite{Lippi2000,Lepri2003,Wang2012}.
In other words, a $1/t$ decay of the heat flux autocorrelation function in a periodic system at equilibrium (in the absence of a temperature gradient), implies a logarithmic divergence of $\kappa$ with system size under non-equilibrium conditions (in the presence of a finite temperature gradient).
Therefore, our predictions can (in principle) be probed experimentally by measuring the size dependence of the thermal conductivity in strained graphene samples.

{\sl Ab initio} LD calculations predict divergence of $\kappa$ in isotropically strained graphene for any amount of applied strain~\cite{Bonini2012}. MD simulations of isotropically strained graphene at $300$ K suggest that $\kappa$ diverges already for $\varepsilon =1\%$, confirming the predictions from LD~\cite{Note1}.
In contrast, uniaxial strain and finite temperature limit divergence to relatively large strain $\varepsilon \geq 2\%$, whereas at lower strain $\kappa$ remains finite. 
Simulations at $T=800$ K confirm the divergence of $\kappa$ along the direction of strain persists for $\varepsilon \geq 2\%$, evidencing no significant difference with respect to the trends observed at room temperature.
In this higher temperature regime, even though far below the Debye temperature of graphene, classical phonon populations approach quantum populations, and quantum effects are mitigated. 
Since phonon lifetimes computed with classical statistics are usually underestimated with respect to those obtained with the correct quantum statistics~\cite{Singh2011}, we can safely argue that the divergent nature of $\kappa$ in strained graphene is not an artifact of classical MD.  

From both analytical models and LD calculations, it appears that the cause for the divergence of $\kappa$ lies in the linearization of long-wavelength ZA modes (Fig.~\ref{fig:dispersions}).
Such alteration of the dispersion relations affects phonon populations, which are probed by computing the vibrational density of states (VDOS) of strained and unstrained samples. Our calculations indeed show a depletion of the VDOS for the ZA modes in the frequency range from $0.1$ THz up to $\approx 15$ THz~\cite{Note1}.
\begin{figure}
\center
\includegraphics[width=0.85\linewidth]{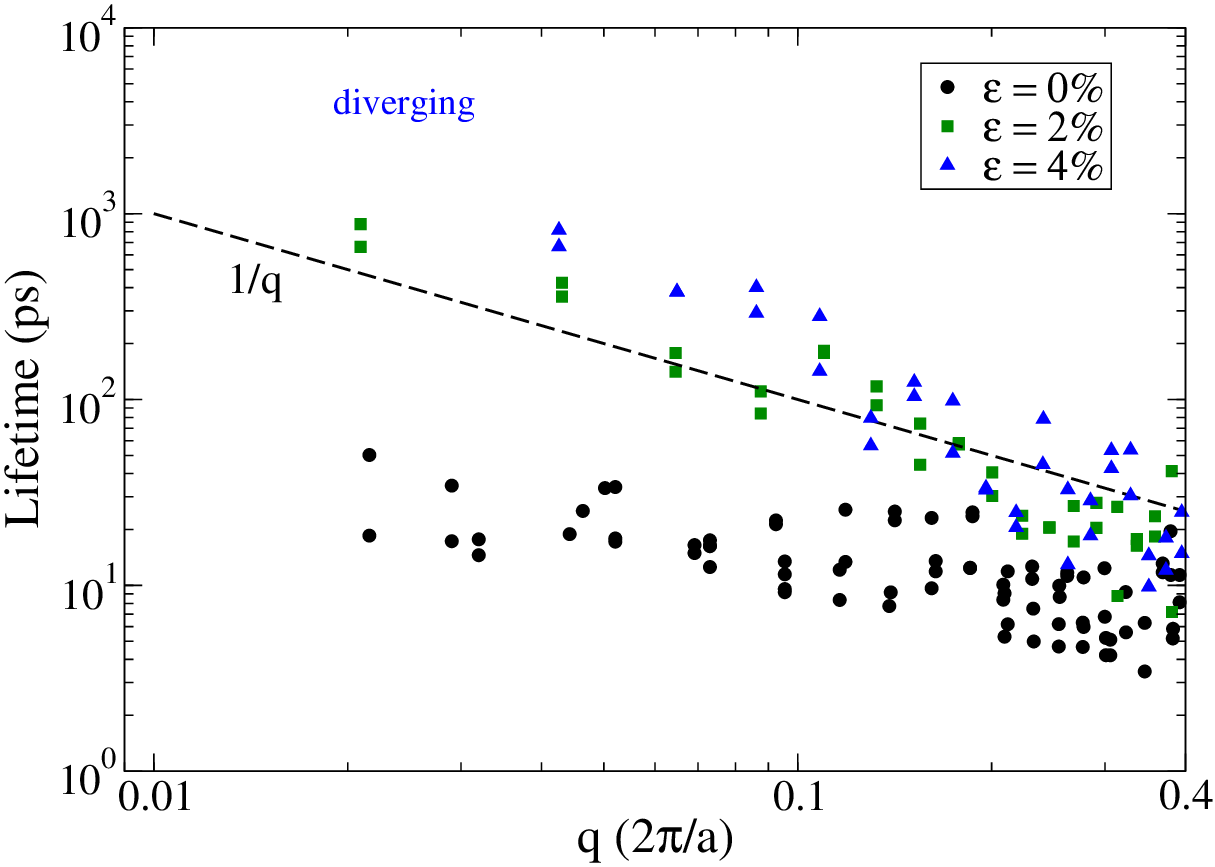}
\caption{(Color online) Lifetimes of the out-of-plane (ZA) phonon modes of suspended graphene model (4400 atoms) with 0, 2 and 4\% uniaxial strain along the armchair direction, at 300 K. For strained samples only the modes propagating along the strain direction ($\Gamma$-M) are shown. Lifetimes at small wavevectors $q$ diverge for the 4\% strained model. The 1/q decay is  shown (dashed line) for reference.}
\label{fig:lifetimes}
\end{figure}
These changes have dramatic effects on phonon lifetimes (Fig.~\ref{fig:lifetimes}), which diverge for large strain and lead to divergence of the thermal conductivity.
Using the solution of the linearized Boltzmann transport equation in the relaxation time approximation, one can express $\kappa$ for a periodic system of finite size as the sum of  the contributions of each phonon mode, as $\kappa = \sum_i^{N_{ph}}  c_i {\rm v}_i^2 \tau_i$. Here $c_i$ is the specific heat of mode $i$, ${\rm v}_i$ its group velocity, and $\tau_i$ its lifetime. 
Since $v_i$ and $c_i$ are always finite, divergence of $\kappa$ implies divergence of $\tau_i$ for some of the modes. 
Phonon lifetimes are computed here as the decay time of the autocorrelation function of the energy of the normal modes in microcanonical MD simulations~\cite{Ladd1986}. 
We indeed observe the presence of ZA  modes with slowly decaying correlation functions when $\varepsilon$ is larger than 2$\%$,  which imply diverging $\tau$.  
The trends of $\tau(\bf{q})$  (Fig.~\ref{fig:lifetimes}) permit the extrapolation of our results to extended systems, for which $\kappa$ is expressed in integral form:
\begin{equation}
\kappa \propto \int_{BZ} dq^2  c(\bf{q}) v^2(\bf{q}) \tau(\bf{q}) \ ,
\label{eq:kappa}
\end{equation}
where the integral is taken over the two-dimensional Brillouin zone (BZ).
The limit for $q \rightarrow 0$ determines whether $\kappa$ diverges. In the classical case $c({\bf q})=k_B/V$ is a constant,  and in general $\tau \propto 1/\omega^\alpha$. The group velocity of the ZA modes along $x$ is ${\rm v}^{ZA}_{x}=d\omega/dq_x$.  For unstrained graphene ${\rm v}^{ZA}_x \propto q_x$, whereas when strain is applied along $x$, ${\rm v}^{ZA}_x$ tends to a constant value.
In unstrained graphene $\kappa$ would diverge for $\alpha \ge 2$, while when strain is applied $\kappa$ diverges for $\alpha\ge 1$.
Our calculations  show that $\alpha$ increases with strain, reaching $\alpha \sim 1$ for $\varepsilon=2\%$, which is consistent with the observed threshold for $\kappa$ divergence.

\subsection{Thermal conductivity of isotopically modified graphene}
\begin{figure}
\center
\includegraphics[width=0.85\linewidth]{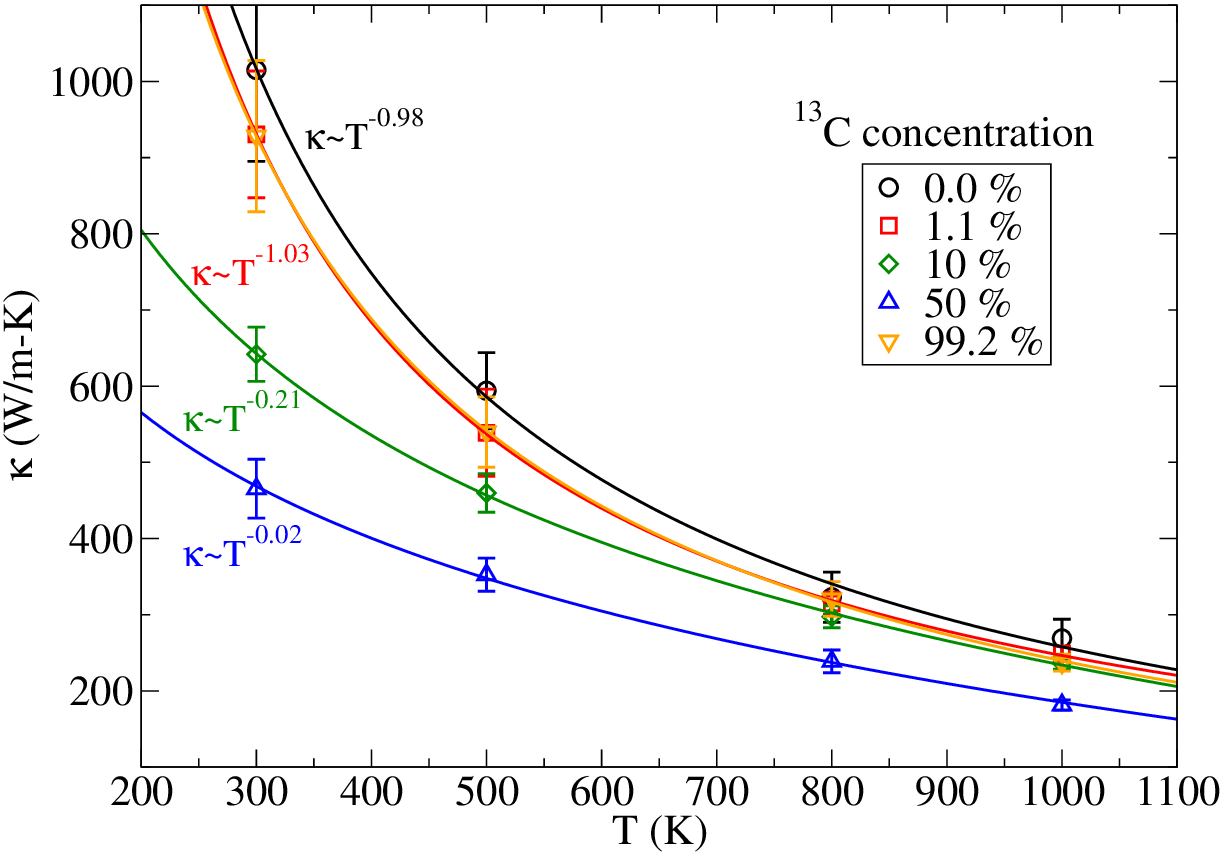}
\caption{(Color online) Temperature dependence of the thermal conductivity of unstrained graphene with different amounts of $^{13}$C. Solid lines are fitted to the data in order to extract the temperature dependence, as reported in the plot.}
\label{fig:kappa.vs.T}
\end{figure}

So far we have presented results for isotopically pure graphene ($100$\% $^{12}$C). Given the demonstration of graphene growth with customized isotopic composition~\cite{Chen2011a}, it is also worth investigating the combined effect of strain and controlled isotopic composition on the thermal conductivity of graphene, checking whether its divergence can be suppressed by mass disorder. 
We consider pure $^{12}$C, natural composition ($1.1\%$ $^{13}$C), and $^{13}$C enriched graphene models (10$\%$, 50$\%$ and 99.2$\%$). In absence of defects the lattice thermal conductivity is limited by phonon-phonon scattering~\cite{Ziman1960}, therefore $\kappa \propto 1/T$. Meanwhile, the scattering of phonons by defects is temperature independent~\cite{Ziman1960}, so that $\kappa(T)$ trends are modified.
In Fig. \ref{fig:kappa.vs.T} we indeed observe $\kappa \sim 1/T$ for isotopically pure graphene, natural graphene and $99.2$\% $^{13}$C isotopically enriched graphene, indicating that for the natural isotopic composition the effect of mass disorder is almost negligible. 
However, as the amount of $^{13}$C increases to 10$\%$ and 50$\%$, $\kappa$ decreases more slowly with $T$ indicating that mass disorder becomes the primary source of phonon scattering. 
This aspect could be exploited in thermal management devices that operate over wide temperature ranges, where it would be undesirable to have large variations in $\kappa$ with $T$.
The ratio between $\kappa$ for isotopically pure and 50$\%$ $^{13}$C-enriched graphene is about 2, and agrees well with recent experimental measurements~\cite{Chen2011a}.

\begin{figure}
\center
\includegraphics[width=0.85\linewidth]{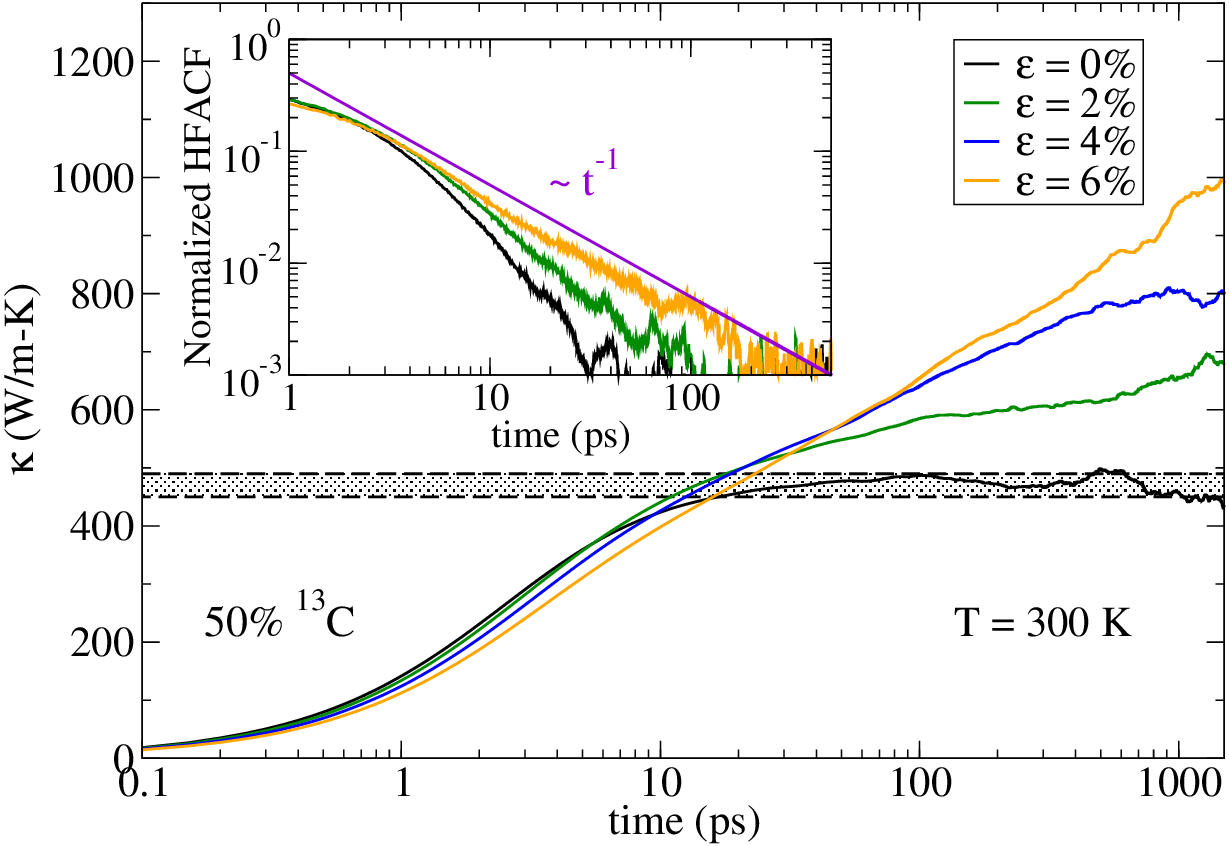}
\caption{(Color online) Thermal conductivity of unstrained and strained isotopically modified graphene at $300$ K, calculated from Eq. (\ref{eq:greenkubo}), with a 50$\%$ concentration of $^{13}$C. The dashed horizontal lines indicate the average $\kappa$ for the unstrained case. The inset shows the  normalized heat flux autocorrelation functions, which decay like $1/t$ for $\varepsilon \geq 2$\%, yielding $\log(t)$ divergence of $\kappa$.}
\label{fig:kappa.vs.timec13}
\end{figure}

Even though large variations of $\kappa$ as a function of isotopic composition are found, when uniaxial tensile strain is applied the general behavior of $\kappa$ for isotopically modified graphene is not qualitatively different from the isotopically pure case. 
$\kappa$ along the strained direction increases for low strain and diverges for $\varepsilon \geq 2\%$ even in samples with the highest isotopic disorder (50$\%$ $^{13}$C), as shown in Fig. \ref{fig:kappa.vs.timec13}. 
The inset shows that above the strain threshold the HFACF decay as $1/t$ and thus $\kappa \sim \log(t)$. 
The ratio between phonon populations shows that in isotopically enriched graphene, as in the isotopically pure case, tensile strain induces similar reductions to the population of ZA modes~\cite{Note1}.
Therefore, mass disorder is not sufficient to suppress the divergence of the lifetimes of low-frequency ZA modes. In fact, these modes have a wavelength of several tens of nm, thus even in the presence of isotopic mass disorder they propagate as in a continuous medium, and are not significantly affected by scattering centers at the atomic scale. 
It is worth noting that mass disorder can not suppress divergence in non-linear models as well~\cite{Lepri2003,Lepri2005}.

\section{Conclusions}

In conclusion, we have shown that heat transport in suspended graphene is controlled by ZA modes, 
which contribute the essential scattering channels to limit the thermal conductivity in unstrained samples. 
In fact, in absence of ZA modes $\kappa$ would diverge with size, as in ideal 2D models.
Uniaxial tensile strain reduces the population of ZA phonon modes at low frequency, makes their zone center group velocity finite and increases their lifetime, thus causing divergence of the thermal conductivity in the strained direction. 
We then predict that $\kappa$ of strained samples would diverge logarithmically with the size of the samples.
It is important to point out that our predictions based on computer simulations are accessible to experiments. 
In an experimental setup, $\kappa$ will always be finite and limited by boundary and defect scattering. Nonetheless, by performing measurements in strained samples of increasing size it should be possible to see a logarithmic dependence of $\kappa$ as a function of size. 
The amount of strain required to observe divergence, $\varepsilon \gtrsim 2\%$, might also be within reach of current experimental techniques.
We also predict that the presence of isotopic mass disorder does not suppress the divergence of $\kappa$ that should be then expected to occur also in samples with natural composition, which are easier to grow than isotopically pure ones.

\acknowledgments{
We are grateful to B. D\"unweg for useful suggestions, to S. Neogi for a critical reading of the manuscript, and to Luciano Colombo for a deep and careful review of our work.
We acknowledge the provision of computational facilities and support provided by Rechenzentrum Garching of the Max Planck society (MPG), and access to the supercomputer JUGENE at the J\"ulich Supercomputing Centre. 
Financial support provided by MPG under the MPRG program.
}

\bibliographystyle{apsrev4-1}

\end{document}